\documentclass[%
 reprint,
superscriptaddress,
 amsmath,amssymb,
 aps, 
noeprint,
]{revtex4-2}

\usepackage{graphicx}
\usepackage{dcolumn}
\usepackage{bm}
\usepackage{hyperref}
\usepackage{graphicx}
\usepackage{amsmath}
\usepackage{caption}
\usepackage{subcaption}
\usepackage{tabularx}
\usepackage{multirow}
\usepackage{siunitx}

\usepackage[mathlines]{lineno}

\begin{document}

\preprint{APS/123-QED}

\title{Bounds on Lorentz Invariance Violation from MAGIC Observation of GRB\,190114C}

%
%
\author{V.~A.~Acciari}
\affiliation{Instituto de Astrof\'{i}sica de Canarias, E-38200 La Laguna, and Universidad de La Laguna, Departamento de Astrof\'{i}sica, E-38206 La Laguna, Tenerife, Spain}
\author{S.~Ansoldi}
\affiliation{Universit\`{a} di Udine, and INFN Trieste, I-33100 Udine, Italy}
\affiliation{Japanese MAGIC Consortium: ICRR, The University of Tokyo, 277-8582 Chiba, Japan; Department of Physics, Kyoto University, 606-8502 Kyoto, Japan; Tokai University, 259-1292 Kanagawa, Japan; RIKEN, 351-0198 Saitama, Japan}
\author{L.~A.~Antonelli}
\affiliation{National Institute for Astrophysics (INAF), I-00136 Rome, Italy}
\author{A.~Arbet Engels}
\affiliation{ETH Zurich, CH-8093 Zurich, Switzerland}
\author{D.~Baack}
\affiliation{Technische Universit\"{a}t Dortmund, D-44221 Dortmund, Germany}
\author{A.~Babi\'c}
\affiliation{Croatian Consortium: University of Rijeka, Department of Physics, 51000 Rijeka; University of Split---FESB, 21000 Split; University of Zagreb---FER, 10000 Zagreb; University of Osijek, 31000 Osijek; Rudjer Boskovic Institute, 10000 Zagreb, Croatia}
\author{B.~Banerjee}
\affiliation{Saha Institute of Nuclear Physics, HBNI, 1/AF Bidhannagar, Salt Lake, Sector-1, Kolkata 700064, India}
\author{U.~Barres de Almeida}
\affiliation{Centro Brasileiro de Pesquisas Físicas (CBPF), 22290-180 URCA, Rio de Janeiro (RJ), Brasil}
\author{J.~A.~Barrio}
\affiliation{IPARCOS Institute and EMFTEL Department, Universidad Complutense de Madrid, E-28040 Madrid, Spain}
\author{J.~Becerra Gonz\'alez}
\affiliation{Instituto de Astrof\'{i}sica de Canarias, E-38200 La Laguna, and Universidad de La Laguna, Departamento de Astrof\'{i}sica, E-38206 La Laguna, Tenerife, Spain}
% ...
\collaboration{MAGIC Collaboration}
\noaffiliation
\author{L.~Nava}
\affiliation{National Institute for Astrophysics (INAF), Osservatorio Astronomico di Brera, 23807 Merate, Italy}
\affiliation{Istituto Nazionale di Fisica Nucleare, Sezione di Trieste, 34149 Trieste, Italy}
\affiliation{Institute for Fundamental Physics of the Universe (IFPU), 34151 Trieste, Italy}

\date{\today}

\begin{abstract}
On January 14, 2019, the Major Atmospheric Gamma Imaging Cherenkov telescopes detected GRB\,190114C above 0.2 TeV, recording the most energetic photons ever observed from a gamma-ray burst. We use this unique observation to probe an energy dependence of the speed of light in vacuo for photons as predicted by several quantum gravity models. 
Based on a set of assumptions on the possible intrinsic spectral and temporal evolution, we obtain competitive lower limits on the quadratic leading order of speed of light modification.
\end{abstract}


\maketitle


\textit{Introduction.---}
Quantum theory and gravity are expected to merge at around the Planck energy (${E_{\mathrm{Pl}} \approx 1.22 \times 10^{19}\,\mathrm{GeV})}$ into a joint, yet unknown theory of quantum gravity (QG). 
Some candidate theories predict a violation or deformation of the Lorentz symmetry, also known as Lorentz invariance violation (LIV, \cite{LIVStringTheory, QGtestsGRB, OpticsQG, NonCommutativeLIV, kappaSystems, WavesNonComm, LoopGraviton}). 
Minuscule effects of LIV could already be visible at energies much lower than $E_{\mathrm{Pl}}$.
One of the manifestations of LIV can be parametrized as energy-dependent corrections to the {\em in vacuo} photon dispersion relation 
\begin{equation}\label{eq:moddispastro}
    E^{2}\simeq
    p^{2}\times\left[1-\sum_{n=1}^{\infty}s\left(\frac{E}{E_{\mathrm{QG,n}}}\right)^{n}\right],
\end{equation}
where $E$ and $p$ are the energy and momentum of the photon, respectively, $E_{\mathrm{QG,n}}$ represents the QG energy scale, 
and $s$ is a theory-dependent factor assuming values $+1$ or $-1$. 
One of the consequences of a modified dispersion relation
is an energy-dependent photon group velocity
\begin{equation}\label{eq:photonvelocity}
    v_\gamma \simeq 1-\sum_{n=1}^{\infty}s \frac{n+1}{2}\left(\frac{E}{E_{\mathrm{QG,n}}}\right)^{n},
\end{equation}
which can be subluminal or superluminal, for $s=+1$ or $s=-1$, respectively. 
This results in an energy-dependent time delay between photons. Taking into account only the leading LIV correction of order $n$, the time delay between photons of energy difference $\Delta E$ is 
\begin{equation}
\Delta t = s  \frac{n+1}{2} D_n(z) \left( \frac{\Delta E}{E_{\mathrm{QG,n}}} \right)^n,
\label{Eq:TimeDelay1}
\end{equation}
where, in setting bounds on LIV, we neglect other potential energy-dependent time delays due to, e.g., the intrinsic emission properties of the source, or massive photons.
A modified dispersion relation would also have an effect on the $\gamma$-$\gamma$ pair-production cross section, and thus on the absorption of $\gamma$ rays \cite{Kifune1999}. However, in this study we focus on investigating effects on the time of flight (TOF) only.
The LIV parameters 
\begin{equation}
    \eta_1 = s \, E_{\mathrm{Pl}}/E_{\mathrm{QG,1}}
    \label{Eq:eta1}
\end{equation} 
and
\begin{equation}
    \eta_2 = 10^{-16} \times s \, E^2_{\mathrm{Pl}}/E^2_{\mathrm{QG,2}},
    \label{Eq:eta2}
\end{equation}
for linear ($n=1$) and quadratic ($n=2$) modification, respectively, are often introduced in Eq. (\ref{Eq:TimeDelay1}) for practicality. 
The information on the comoving distance between the source and the detector is included in $D_n(z)$ \cite{Jacob2008a}
\begin{equation}
D_n(z) = \frac{1}{H_0} \int_0^z  \frac{(1+\zeta)^n}{\sqrt{\Omega_{\Lambda} + (1+\zeta)^3 \Omega_m}} d\zeta,
\end{equation}
where $\Omega_{\Lambda}$, $H_0$, and $\Omega_m $ denote the cosmological constant, the Hubble parameter and the matter fraction, respectively. In this Letter, we use $H_0 = 70  \, \mathrm{km} \, \mathrm{s}^{-1}  \, \mathrm{Mpc}^{-1}$, $\Omega_{\Lambda} = 0.7$, and $\Omega_m = 0.3$. 
The systematic effect introduced by these relatively coarse values and their variations is negligible compared to the sensitivity of our analysis. 

To date, the most stringent lower limits on the QG energy scale, resulting from TOF studies, were set using the observation of GRB\,090510 with the Large Area Telescope (LAT) on board the \textit{Fermi} satellite for the linear case, and observations of active galactic nucleus Mrk\,501 with the H.E.S.S. telescopes for the quadratic case. 
The values for the subluminal (superluminal) scenario are ${E_{\mathrm{QG,1}} > 2.2\times10^{19}\,\mathrm{GeV}}$ (${E_{\mathrm{QG,1}} > 3.9\times~10^{19}\,\mathrm{GeV}}$) \cite{Vasileiou2013} (although the analysis reported in \cite{Ellis2019} does not support this limit) and ${E_{\mathrm{QG,2}} > 8.5 \times 10^{10}\,\mathrm{GeV}}$ (${E_{\mathrm{QG,2}} > 7.3 \times 10^{10}\,\mathrm{GeV}}$) \cite{Abdalla2019}. 
A third class of sources used for the TOF studies on $\gamma$ rays are pulsars. Results obtained on Crab pulsar observations with the Major Atmospheric Gamma Imaging Cherenkov (MAGIC) telescopes can be found in \cite{Ahnen2017}.\\
\indent A potential LIV-induced time delay increases with the distance of the source and the energy of the photons. The sensitivity to detect the TOF effect depends inversely on the timescale of the signal variability, which provides a time reference with respect to which time delays can be measured.
Gamma-ray bursts (GRBs) are among the most distant $\gamma$-rays sources and their signal varies on subsecond timescales. As such, they were identified as excellent candidates for LIV studies many years ago \cite{QGtestsGRB} and already detected frequently in the high energy (HE, $E \lesssim 100$\,GeV) regime with detectors on board the \textit{Fermi} satellite \cite{Ajello2019_short}. However, they are notoriously difficult to detect in the very high energy (VHE, $E>100$\,GeV) band. The recent detection of GRB\,190114C at redshift $z=0.4245 \pm 0.0005$ \cite{Selsing2019, Castro-Tirado2019} with the MAGIC telescopes was the first one reported at TeV energies \cite{GRB190114C_MAGICdiscovery}.\\
\indent In this Letter, we present the results of a LIV study based on the VHE $\gamma$-ray signal from GRB\,190114C. 
The MAGIC observations and data analysis are presented in the next section. 
The TOF analysis method is described in the maximum likelihood analysis section.
Then, we present our results and discuss differences between methods. The most important conclusions are summarized in the final section.

\textit{MAGIC observation of GRB\,190114C.---}
MAGIC is a system of two 17-meter-diameter imaging atmospheric Cherenkov telescopes \cite{Aleksic2016a_short}. Thanks to their relatively light weight and fast repointing capability, the MAGIC telescopes are optimally designed to investigate GRBs as one of their primary goals. They are located in the Roque de los Muchachos observatory on the Canary Island of La Palma at about 2200 meters above the sea level. \\
\indent The MAGIC telescopes detected a strong VHE $\gamma$-ray signal from GRB\,190114C \cite{GRB190114C_MAGICdiscovery,GRB190114C_MWL}, after the initial trigger on January 14, 2019 at 20:57:03 universal time (hereafter $T_0$).
The intrinsic spectrum averaged over the time window from $T_0+62$ seconds to $T_0+2400$ seconds is well fitted with a power law function with index $\alpha = -2.5 \pm 0.2$ \cite{GRB190114C_MWL}, 
and it appears to be constant for the duration of the observation. 
The intrinsic integrated flux in the energy range $0.3 - 1$\,TeV decays as a power law with time decay index $\beta = -1.51 \pm 0.04$ \cite{GRB190114C_MWL}. This observation includes the highest energy photons ever detected from a GRB. 
For our LIV analysis, we selected events recorded during the first 19 minutes of observation of  GRB\,190114C, with stable observational conditions and covering approximately $90\%$ of all observed events. 
The signal events were extracted from the so-called ON region, a circular sky region of radius $0.1^\circ$--$0.2^\circ$ (depending on the energy) around the position of the source, which also contains background events. The background content of the ON region was estimated counting events in three simultaneous OFF regions within the field of view, and of the same size as the ON region. This resulted in a total of $N_{\mathrm{ON}}=726$ and $N_{\mathrm{OFF}}=119$ events (i.e., ${119/3=39.67}$ estimated background events in the ON region), with estimated energies from $E_{\mathrm{min}} = 300$\,GeV to $E_{\mathrm{max}} = 1955$\,GeV and arrival times from $t_{\mathrm{min}} = 62$\,s to $t_{\mathrm{max}} =1212$\,s after $T_0$.

\textit{Maximum likelihood analysis.---}
We estimate the value of the LIV parameters $\eta_n$ (${n \in \{1,2\}}$) using the maximum likelihood method; first employed in TOF studies of LIV using Cherenkov telescopes in \cite{Martinez2009}. 
This method allows us to search for optimal value of $\eta_n$, while taking into account source-intrinsic temporal and energy distributions of events, as well as our instrument's response functions. 
First we define the probability distribution function (PDF) for a signal event. It gives us the probability of detecting a photon of estimated energy $E_{\mathrm{est}}$ at time $t$ as
\begin{multline}
f_s( t, E_{\mathrm{est}}  \, | \,  \eta_n, I ) \propto \\
   \int_0^{\infty} dE \;  \Phi_1 [t - \Delta t(E, \eta_n)] \, \Phi_2 (E) \\ \times \, F(E)  \, A_{\mathrm{eff}}(E) \, G \left( E_{\mathrm{est}}, E \right),
\label{Eq:PDF}
\end{multline}
where $\Phi_1 [t - \Delta t(E, \eta_n)]$ represents the temporal distribution of $\gamma$ rays (modified for the potential LIV-induced time delay), and $\Phi_2(E)$ is the energy distribution of $\gamma$ rays at the source. VHE $\gamma$ rays are partially absorbed by the extragalactic background light (EBL), resulting in the observed spectrum being softer compared to the intrinsic one. $F(E)$ is the EBL attenuation, which in this Letter we computed using the model of A. Dom\'inguez \textit{et al.} \cite{Dominguez2011} with $z=0.4245$. $A_{\mathrm{eff}}(E)$ is the acceptance of our instrument, i.e., the probability of detecting a photon of energy $E$. $G \left( E_{\mathrm{est}}, E \right)$ accounts for the finite energy resolution of our instrument. It is the PDF for the true energy $E$ of a photon to be measured as $E_{\mathrm{est}}$. The parameters of the intrinsic energy and temporal photon distributions are represented with $I$ and treated as nuisance parameters.

Because there is no evidence of change of intrinsic spectrum with time \cite{GRB190114C_MAGICdiscovery}, we assume that the intrinsic energy and temporal distributions are mutually independent (a systematic effect introduced with this assumption is investigated in the end of the results section). The intrinsic energy distribution is modeled with a power law as described in the previous section. 
$A_{\mathrm{eff}}(E) $ and $G \left( E_{\mathrm{est}}, E \right)$ are obtained from Monte Carlo simulations. \\ 
\indent At this point we need to define a functional form for $\Phi_1 [t - \Delta t(E, \eta_n)]$. As described in the previous section, the light curve measured by MAGIC is a monotonic and smooth power law. A back-of-the-envelope calculation can show that taking a power law temporal distribution and applying an energy-dependent time delay will result again in a power law temporal distribution (with different parameters), and any effect of an energy-dependent time delay would be impossible to detect. 
\begin{figure}[!t]
 \includegraphics[width=\linewidth]{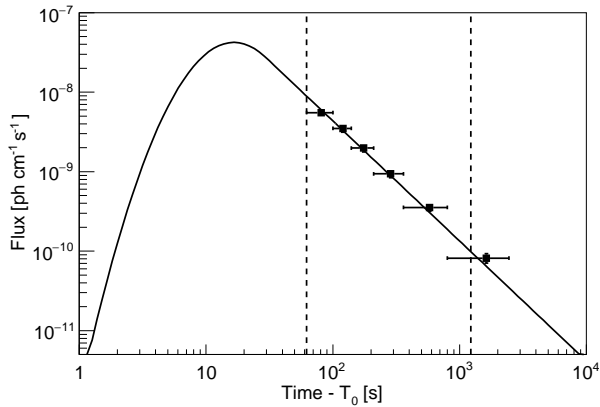}
 \caption{Intrinsic LC model. The points represent the $\gamma$-ray flux measured by MAGIC in the 0.3--1\,TeV energy range, while the full line represents the LC model reported in \cite{GRB190114C_MWL}. The vertical dashed lines represent the bounds of the time interval considered in our analysis.}
 \label{fig:LC}
\end{figure}
This inability to set strong constraints on the emission time  poses the main limitation to our analysis sensitivity. Therefore, we need to make an assumption on the shape of the intrinsic temporal distribution of $\gamma$ rays beyond the interval of MAGIC observations. For this, we adopt the following two approaches:
(1) in the \textit{minimal} approach our only assumption is that the $\gamma$-ray emission started at $T_0$, and we avoid making any further assumptions about the temporal distribution of the photons. Therefore, we define the time model as a step function:
    \begin{equation}
     \Phi_1 (t)  = 
     \begin{cases}
      0 \quad & t < T_0, \\
      k \quad & t \geq T_0,
     \end{cases}
     \label{eq:LCstep}
    \end{equation}
    where $k$ is an arbitrary constant absorbed in the PDF normalization.
    Any event has equal probability of being emitted at any time after $T_0$ (and 0 probability of being emitted before $T_0$), thus avoiding any assumption about the intrinsic temporal photon distribution. In this sense the approach is conservative, since the only assumption is that there was no $\gamma$-ray emission before $T_0$. 
(2) In the \textit{theoretical} approach, we adopt the intrinsic temporal distribution from \cite{GRB190114C_MWL}. The temporal evolution of the afterglow forward shock emission in the $0.3 - 1$\,TeV energy range was modeled based on multiwavelength (MWL) observations and theoretical considerations. The light curve (LC) model is shown in Fig.~\ref{fig:LC}. For the purposes of this study, we parametrized the LC for the duration of the observation as follows:
\begin{equation}
 \Phi_1 (t)  \propto 
 \begin{cases}
 0 \quad & t < T_0 \equiv 0 \\
h(t) \quad & T_0< t < T_1 \\
h(T_1) \, (t/T_1)^{\beta} \quad & t > T_1
 \end{cases}
 \label{eq:LC}
\end{equation}
where $ h(t) = t^{7.3 - 1.3 \ln(t)}$ and $T_1 = 30$\,s
\cite{GRB190114C_MWL}. 
In both approaches, all 726 events from the ON region are used for the likelihood maximization. The intrinsic parameters $\alpha$ and $\beta$ are treated as nuisance parameters, the latter one being only applicable for the theoretical approach. 

Finally, the likelihood function can be written as
\begin{widetext}
\begin{align}
\begin{split}
  \mathcal{L} & \left( \eta_n ; \;  I  \; | \; \{ t^{(i)}, E_{\mathrm{est}}^{(i)} \}_{i=1,...,N_{\mathrm{ON}}} \, , \; N_{\mathrm{ON}}, N_{\mathrm{OFF}} \right)  \;  = \; P(I) \\
& \times \prod_i^{N_{\mathrm{ON}}}   \left(
 \frac{N_{\mathrm{ON}}-N_{\mathrm{OFF}}/\tau}{N_{\mathrm{ON}}} \, \frac{ f_s( t^{(i)}, E_{\mathrm{est}}^{(i)}  \, | \, \eta_n , I ) }{   \int_{E_\mathrm{min}}^{E_\mathrm{max}}dE_{\mathrm{est}} \int_{t_\mathrm{min}}^{t_\mathrm{max}}dt \;  f_s( t, E_{\mathrm{est}}  \, | \, \eta_n ,  I ) }  \; + \;  
 \frac{N_{\mathrm{OFF}}}{\tau N_{\mathrm{ON}}} \, \frac{f_b(t^{(i)}, E_{\mathrm{est}}^{(i)})}{  \int_{E_\mathrm{min}}^{E_\mathrm{max}}dE_{\mathrm{est}} \int_{t_\mathrm{min}}^{t_\mathrm{max}}dt \;  f_b(t, E_{\mathrm{est}} ) } 
\right),
\end{split}
\label{Eq:Likelihood}
\end{align}
\end{widetext}
where $E_{\mathrm{est}}^{(i)}$ and $t^{(i)}$ are the estimated energy and arrival time, respectively, of event $i$. $P(I)$ is the PDF of the parameters describing the intrinsic energy and temporal evolution of the source;
for the theoretical approach, we assume that $\alpha$ and $\beta$ are distributed according to normal distributions centered, respectively, at $-2.5$ and $-1.51$, with standard deviations 0.2 and 0.04, respectively \cite{GRB190114C_MWL}. 
$\tau$ is the ratio of exposure time between the background and the signal regions. In our case $\tau=3$ (see the previous section). 
The background PDF $f_b(t, E_{\mathrm{est}})$ is obtained assuming a uniform distribution in time (justified by the stable observation conditions), while for estimating the energy distribution we use events collected with MAGIC when pointing under the same observational conditions to regions of the sky with no known $\gamma$-ray sources. 

We then compute
\begin{equation}
    L = -2 \ln{ \left( \frac{ \text{max}(\mathcal{L})_{I} }{\text{max}(\mathcal{L})_{\eta_n, I}} \right) }
    \label{Eq:DeltaL}
\end{equation} 
as a function of $\eta_n$, and search for $\eta_n$ which minimizes $L$. 
In Eq.~(\ref{Eq:DeltaL}) we have introduced the notation $\text{max}(\mathcal{L})_{I} \equiv \mathcal{L}(x,\hat{I})$ where $\hat{I}$ maximizes $\mathcal{L}$ for a given value of $x$. 
In this way we treat all the intrinsic parameters in the maximum likelihood as nuisance parameters. 
This approach has the advantage that 
uncertainties 
on the intrinsic properties of the source (namely the spectral index $\alpha$ and the time index $\beta$ of the integral flux power-law decay defined in Eq.~\ref{eq:LC}) are included in the obtained confidence intervals (CIs) for the QG energy scale.

\begin{figure}[!t]
\centering
\includegraphics[width=\columnwidth]{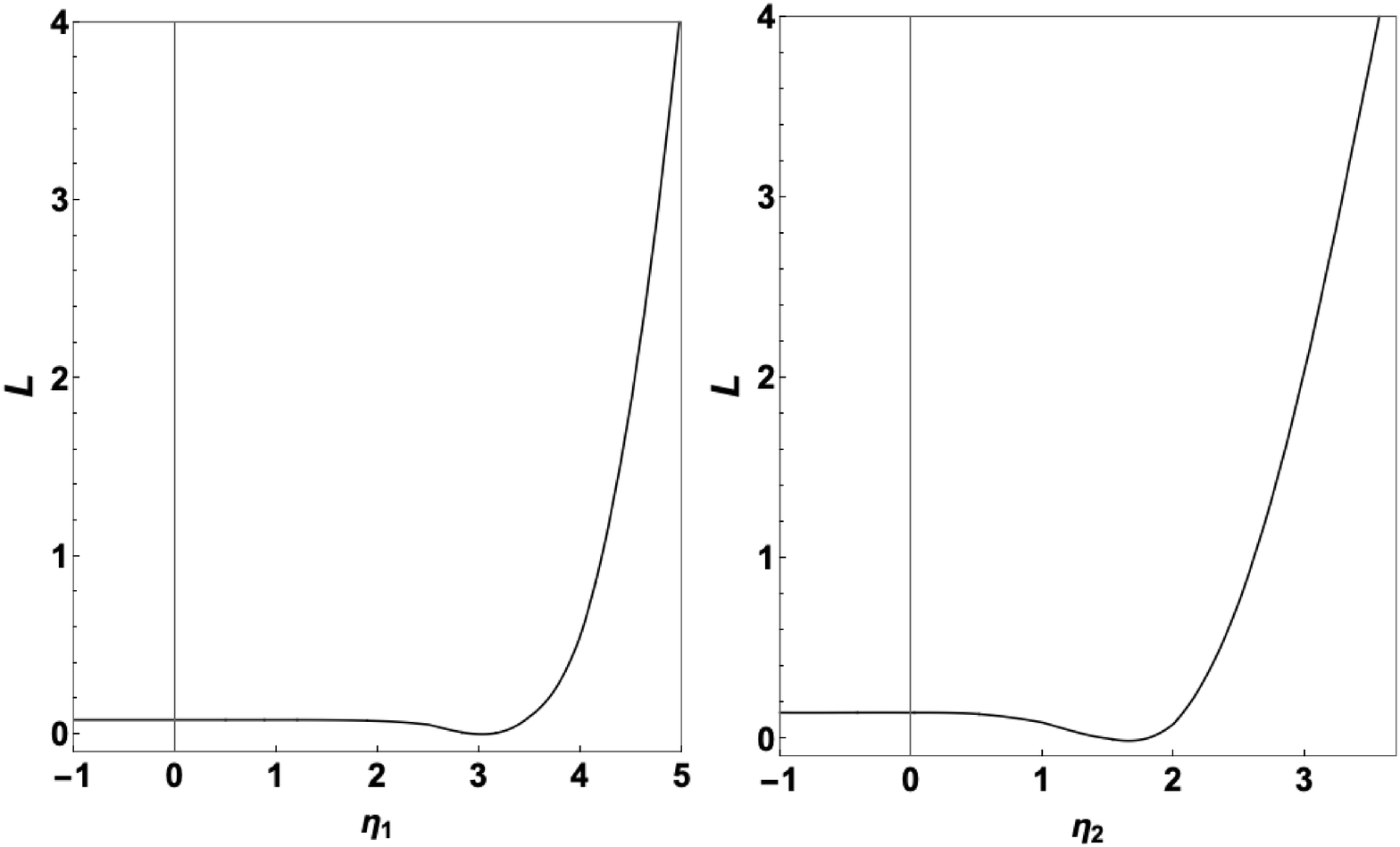}
\caption{Likelihood profile for the linear (left) and quadratic (right) case, using the minimal model for the intrinsic LC.}
\label{fig:Agnostic_LC_likelihood}
\end{figure}

\begin{figure}[!t]
\centering
\includegraphics[width=\columnwidth]{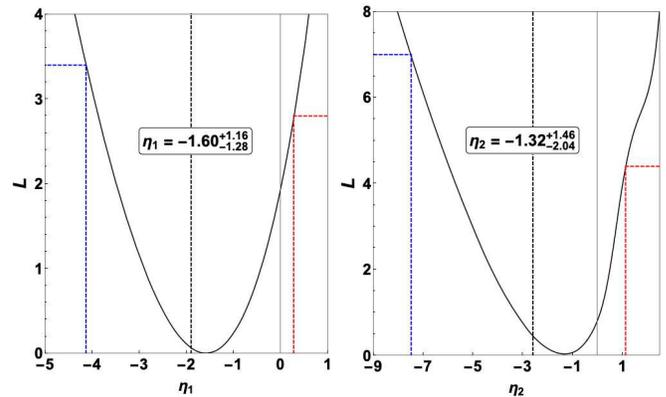}
\caption{Likelihood profile for the linear (left) and quadratic (right) case, using the theoretical model for the intrinsic LC. The black dashed line represents the bias obtained from mock data sets (see Section~A of the Supplemental Material \cite{Supplemental}). 
The point at which the likelihood is equal to the calibrated $95 \%$ CIs  is instead shown using blue and red dashed lines for the lower and upper limit, respectively (see Section~B of the Supplemental Material \cite{Supplemental}).}
\label{fig:MWL_LC_likelihood}
\end{figure}

\textit{Results and discussion.---}
We first perform our analysis adopting minimal model for the intrinsic light curve [Eq.~(\ref{eq:LCstep})]. The results for $L$ vs $\eta_n$ for the linear and quadratic modification are shown in Fig.~\ref{fig:Agnostic_LC_likelihood}.
As expected, the likelihood profiles are constant (despite small fluctuation) and minimal for negative and small values of $\eta_n$, which correspond to superluminal or mild subluminal behavior. In this model all photons have equal probability of being emitted at any time after $T_0$. Therefore, the value of $L$ will not change as long as the time delay implies all photons were indeed emitted after $T_0$. Once the $\eta_n$ becomes positive enough ($\eta_1\simeq 3.5$ for the linear, and $\eta_2\simeq 2.1$ for the quadratic modification), implying stronger subluminal behavior, the time delay will imply some photons should have been emitted before $T_0$. For instance, for the linear case and $\eta_1=5$, we expect a delay of $\sim 83$\,s for $\gamma$ rays of $E=1$\,TeV, whereas we have observed an $E_{\mathrm{est}}=1.07$ TeV event at $t=T_0+73.6$\,s, meaning it should have been emitted before $T_0$. These photons do not contribute to the likelihood function any more, and the likelihood values rapidly decrease. 
Note that, since $L$ has no strict minimal value and it is constant for negative and small positive values of $\eta$, the minimal approach can only be used to set upper limits on the value of $\eta$. This corresponds to setting lower limits on QG energy scale for subluminal behavior.\\ 
\indent Before performing the analysis using the theoretical model for the intrinsic light curve [Eq.~(\ref{eq:LC})], we study the sensitivity and influence of systematic effects on this approach. For that purpose, we perform analysis on 1000 LIV-free mock data sets, from which we estimate the bias associated to the maximum likelihood analysis applied to this particular temporal and energy distributions (see Section~A of the Supplemental Material \cite{Supplemental} for details).
From the distribution of the results on mock data sets, we find that our analysis has a bias towards negative values of LIV parameter $\eta$. In particular, we obtain ${\eta_{1,\mathrm{bias}} = -1.9}$ and ${\eta_{2,\mathrm{bias}} = -2.6}$. 
Analysing the real data and using the theoretical model for the intrinsic light curve (Eq.~\ref{eq:LC}), we find that the likelihood is maximal for $\eta_1=-1.6$ and $\eta_2=-1.32$ (see Fig.~\ref{fig:MWL_LC_likelihood}) in the linear and quadratic case, respectively. We correct these values for the bias to get the best fit values $(\eta^{\mathrm{BF}})$ reported in Table~\ref{Tab:results}. 
Our results appear to be consistent with the null hypothesis $(\eta=0)$ (see Section~A of the Supplemental Material \cite{Supplemental}), i.e., no energy-dependent time delay, or $E_{\mathrm{QG}}\rightarrow\infty$. 
Therefore, we set upper limits on $\eta$ by constructing calibrated $95 \%$ CIs from the reshuffled-bootstrapped samples. The procedure adopted from \cite{Vasileiou2013} is described in Section~B of the Supplemental Material \cite{Supplemental}.
The obtained calibrated CIs are reported in Table~\ref{Tab:results} (assuming $\eta_{\mathrm{bias}}=0$ in the case of  the minimal LC).
Finally, using Eqs.~(\ref{Eq:eta1}) and (\ref{Eq:eta2}), these values are translated into the limits on the energy scale $E_{\mathrm{QG}}$ at $95 \%$ confidence level and reported in Table \ref{Tab:results}.
\begin{table}[htbp]
\caption{Values of the $95 \%$ lower (LL) and upper (UL) limits and the best fits (BF) obtained for $\eta_n$ using the theoretical intrinsic LC model, after applying bias correction and CI calibration. Only upper limits can be set with the minimal approach (see text). Values are reported for the linear ($n=1$) and quadratic ($n=2$) cases.}
\begin{ruledtabular}
\begin{tabularx}{\textwidth}{c | c | r c l }
LC & Minimal & \multicolumn{3}{c}{\multirow{ 2}{*}{Theoretical (\cite{GRB190114C_MWL})}} \\
 model & (step function) & \\
 \colrule
 & $\eta^{\mathrm{UL}}$ & $\eta^{\mathrm{LL}}$ & $\eta^{\mathrm{BF}}$ & $\eta^{\mathrm{UL}}$ \\ \colrule
$\eta_1$ & 4.4 & -2.2 & 0.3 & 2.1 \\
$\eta_2$ & 2.8 & -4.8 & 1.3  & 3.7 \\ \colrule\colrule
 & subl. & superl. & & subl. \\ \colrule
$E_{\mathrm{QG,1}}$\,[$10^{19}$\,GeV] & 0.28 & 0.55 & & 0.58 \\
$E_{\mathrm{QG,2}}$\,[$10^{10}$\,GeV] & 7.3 & 5.6 & & 6.3 \\
\end{tabularx}
\end{ruledtabular}
\label{Tab:results}
\end{table}\\
\indent A possible change of spectral index of GRB\,190114C with time was reported in \cite{GRB190114C_MWL}. We investigated the resulting systematic effect on $\eta$, and found that it is less than $5 \%$ in all cases. 
Additionally, using a dedicated study with Monte Carlo simulations, we computed that the limits would degrade by up to $18 \%$ ($29 \%$) in subluminal (superluminal) case, should the Cherenkov light collected by the telescopes be overestimated by $15 \%$ in our analysis, which is a conservative assumption.

\textit{Conclusions and summary.---}
MAGIC discovered a $\gamma$-ray signal above 0.2\,TeV from GRB\,190114C, detecting the highest energy photons from a GRB. Using conservative assumptions on the intrinsic spectral and temporal emission properties, we searched for an energy-dependent delay in arrival time of the most energetic photons, testing in vacuo dispersion relations of VHE photons. We assumed two different models for the LC: minimal and theoretical, described in detail in the maximum likelihood analysis section. In both cases, our results are compatible with the null hypothesis of no time delay. 
We set lower limits on LIV energy scale. Our results for the linear modification of the photon dispersion relation
${E_{\mathrm{QG,1}} > 0.58\times10^{19}\,\mathrm{GeV}}$ 
(${E_{\mathrm{QG,1}} > 0.55\times10^{19}\,\mathrm{GeV}}$) 
for the subluminal (superluminal) case are approximately a factor 4 (7) below the most constraining lower limits on $E_{\mathrm{QG,1}}$ obtained from TOF method on GRB\,090510 \cite{Vasileiou2013}. This is expected because of a significantly larger distance of GRB\,090510 ($z=0.9$, compared to 0.4245 of GRB\,190114C), as well as a shorter variability timescale, since \textit{Fermi}-LAT observations of GRB\,090510 include a full coverage of the emission.
In the quadratic case, the analysis is more sensitive to the highest photon energies in the data sample (estimated $E_{\mathrm{max}} = 1955$\,GeV, compared to $E_{\mathrm{max}} = 31$\,GeV for GRB\,090510 \cite{Vasileiou2013}).
As a result, our lower limits on the energy scale ${E_{\mathrm{QG,2}} > 6.3 \times 10^{10}\,\mathrm{GeV}}$ (${E_{\mathrm{QG,2}} > 5.6 \times 10^{10}\,\mathrm{GeV}}$) for the subluminal (superluminal) case are more constraining than the ones in \cite{Vasileiou2013}. At the same time, our results are comparable to the ones from \cite{Abdalla2019}. GRB\,190114C is at redshift more than one order of magnitude higher than Mrk\,501; however, the measured spectrum of Mrk\,501 reaches an order of magnitude higher energies \cite{Abdalla2019}, resulting in comparable sensitivities.
It is worth noting that MAGIC observed a featureless afterglow phase of the GRB\,190114C, limiting the sensitivity of our LIV analysis. We are looking forward to VHE observations of an expectedly feature-rich GRB prompt phase, which would enhance the analysis sensitivity to LIV effects.

\begin{acknowledgments}
%
%
We would like to thank the Instituto de Astrof\'{\i}sica de Canarias for the excellent working conditions at the Observatorio del Roque de los Muchachos in La Palma. The financial support of the German BMBF and MPG, the Italian INFN and INAF, the Swiss National Fund SNF, the ERDF under the Spanish MINECO (Grants No. FPA2017-87859-P, No. FPA2017-85668-P, No. FPA2017-82729-C6-2-R, No. FPA2017-82729-C6-6-R, No. FPA2017-82729-C6-5-R, No. AYA2015-71042-P, No. AYA2016-76012-C3-1-P, No. ESP2017-87055-C2-2-P, and No. FPA2017‐90566‐REDC), the Indian Department of Atomic Energy, the Japanese JSPS and MEXT, the Bulgarian Ministry of Education and Science, National RI Roadmap Project No. DO1-153/28.08.2018 and the Academy of Finland Grant No. 320045 is gratefully acknowledged. This work was also supported by the Spanish Centro de Excelencia ``Severo Ochoa'' SEV-2016-0588 and SEV-2015-0548, and Unidad de Excelencia ``Mar\'{\i}a de Maeztu'' MDM-2014-0369, by the Croatian Science Foundation (HrZZ) Project No. IP-2016-06-9782 and the University of Rijeka Project No. 13.12.1.3.02, by the DFG Collaborative Research Centers SFB823/C4 and SFB876/C3, the Polish National Research Centre Grant No. UMO-2016/22/M/ST9/00382 and by the Brazilian MCTIC, CNPq, and FAPERJ.
This project has received funding from the Foundation Blanceflor Boncompagni Ludovisi, n\'ee Bildt.
This project has received funding from the European Union's Horizon 2020 research and innovation programme under the Marie Sk\l{}odowska-Curie Grant Agreement No. 754510.
L. Nava acknowledges funding from the European Union's Horizon 2020 Research and Innovation programme under the Marie Sk\l{}odowska-Curie Grant Agreement No. 664931.
The authors would like to acknowledge networking support by the COST Action CA18108.
\end{acknowledgments}

\bibliography{main}

\onecolumngrid
\noindent\rule{17.5cm}{0.3pt}
\section*{Supplemental Material\\ for\\ Bounds on Lorentz invariance violation from MAGIC observation of GRB\,190114C}
\begin{center} 
The MAGIC Collaboration
\end{center}
\noindent\rule{17.5cm}{0.3pt}
\setcounter{page}{1}
\setcounter{figure}{0}
\setcounter{equation}{0}

\subsection{\label{sec:AppBias}Analysis sensitivity}
\begin{figure}[!h]
\centering

\begin{minipage}{.5\textwidth}
\centering
\includegraphics[width=0.9\linewidth]{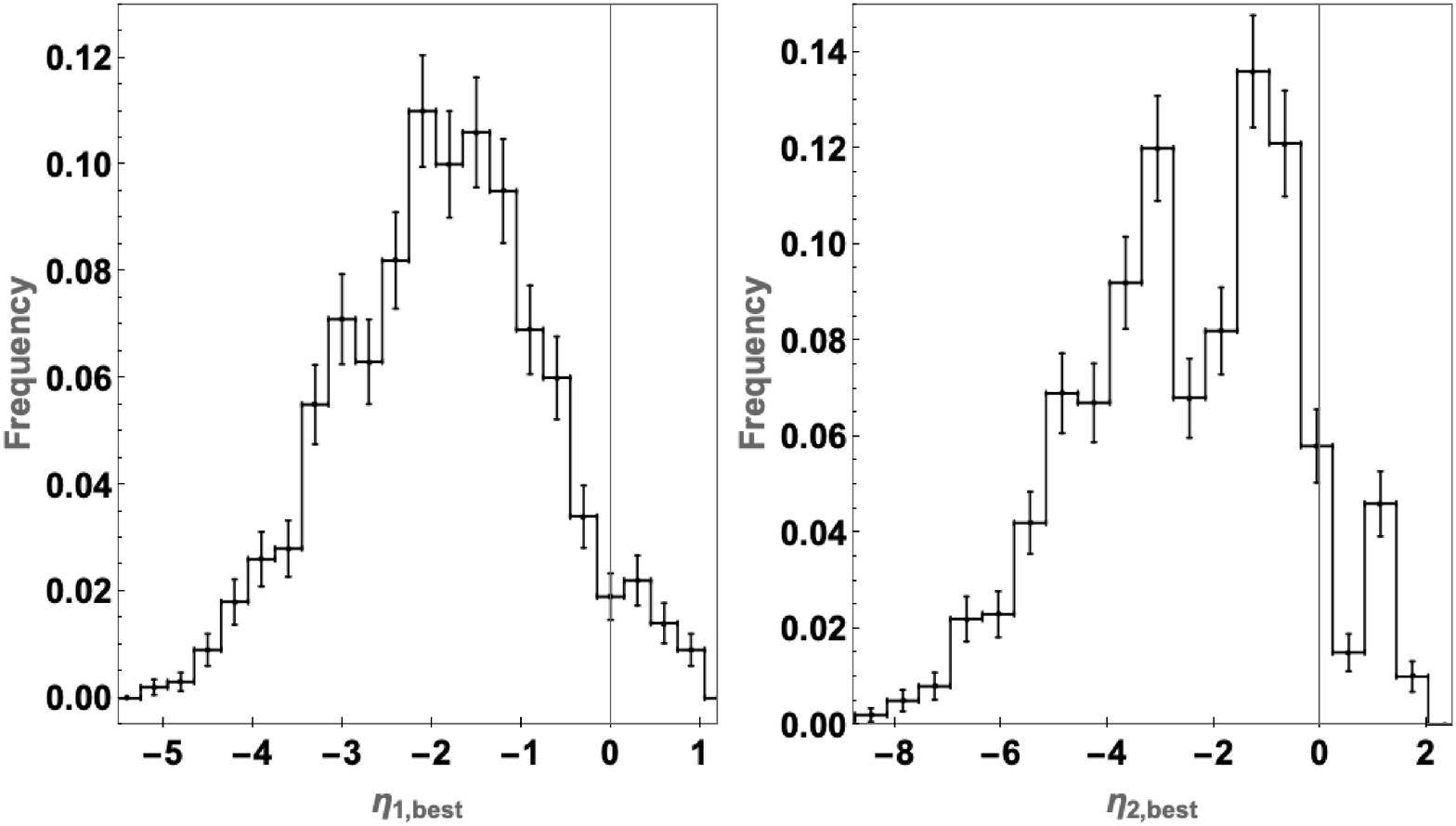}
\caption{Distribution of best fits of $\eta_1$ (linear case, left) and $\eta_2$ (quadratic case, right), obtained from reshuffled-bootstrapped samples and using the theoretical assumption for the intrinsic LC.}
\label{fig:BV_distribution}
\end{minipage}%
\begin{minipage}{.5\textwidth}
\centering
\includegraphics[width=0.9\linewidth]{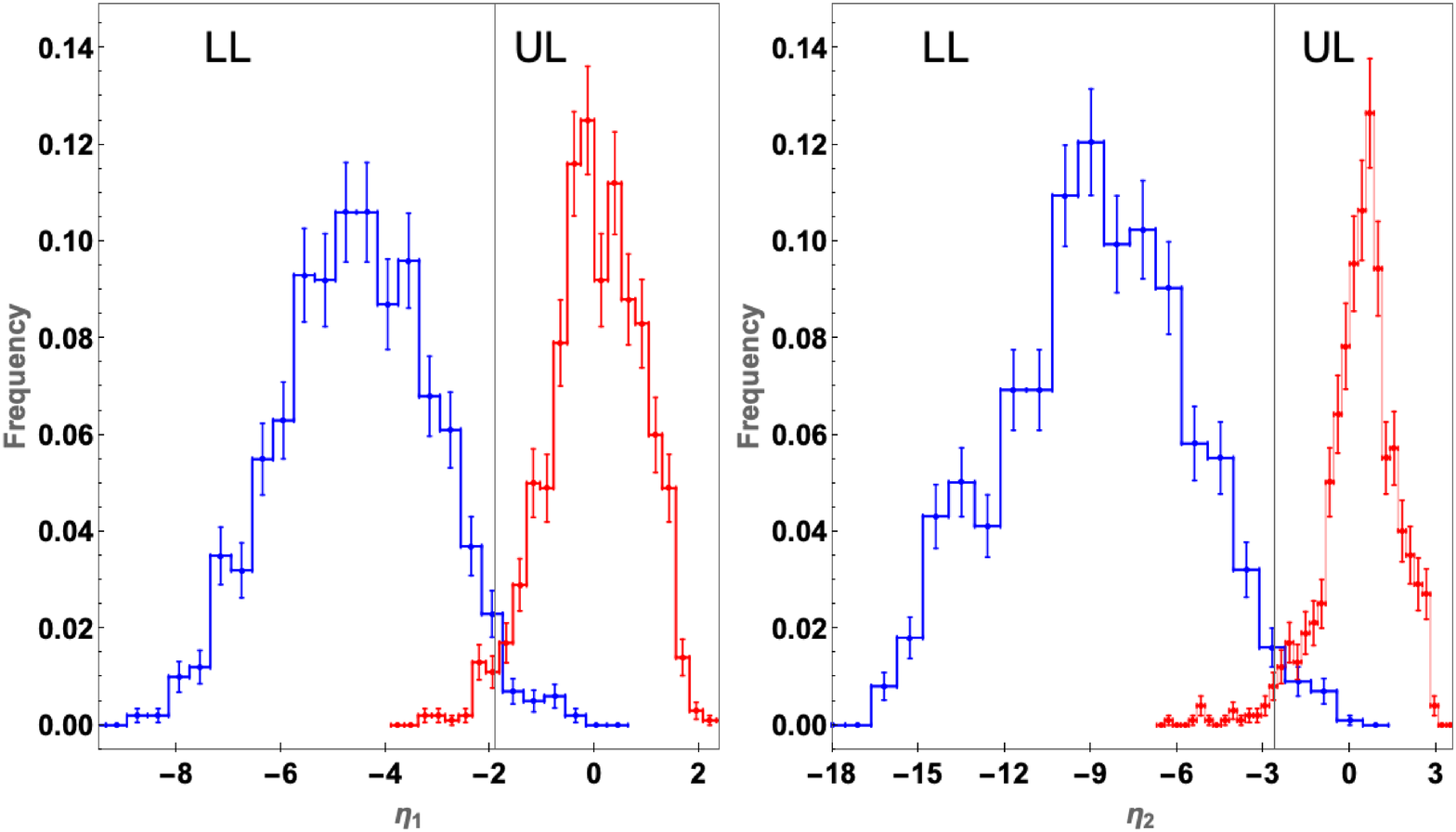}
\caption{Distribution of lower (blue) and upper limits (red) for the linear (left) and quadratic case (right), obtained from reshuffled-bootstrapped samples and using the theoretical assumption for the intrinsic LC. The vertical lines indicate respective bias values.}
\label{fig:Limits_distribution}
\end{minipage}
\end{figure}

When building the likelihood function, we have assumed certain spectral and temporal distributions of the signal events. These were determined using our data, however, only up to a certain precision and under several theoretical assumptions \cite{GRB190114C_MAGICdiscovery,GRB190114C_MWL}. Therefore, we cannot determine in an unbiased way their level of accuracy. In particular, we know that the minimal model of the intrinsic light curve (Eq.~(\ref{eq:LCstep}) in the main text) does not correctly describe the temporal distribution of the signal events. Because of this, we cannot presume that the PDF for $L$ is a $\chi^2$ with one degree of freedom, or that the estimator of $\eta$ distributes as a Gaussian around the true value $\eta_\mathrm{true}$. Thus, to evaluate the PDF of the $\eta$ estimator we apply the maximum likelihood method to 1000 mock data sets. Each of them is generated starting from the measured data set, first ``reshuffling'' the event arrival times, and then applying once the bootstrapping resampling technique. Reshuffling consists of reassigning randomly the measured arrival times to the different observed events. In this way, we remove any energy-time correlation present in the data (in particular, any LIV effect), without altering the overall spectral and temporal distributions of the signal. Bootstrapping creates samples of the same size by randomly selecting events (repetition is allowed) from the reshuffled data set, and therefore allows the measured spectral and temporal distributions to vary within their natural statistical uncertainties.\\
\indent We maximize the likelihood for each of the reshuffled-bootstrapped samples, and make the histogram of the resulting best fits. This gives us the PDF of our estimator, shown in Fig.~\ref{fig:BV_distribution}. Since the reshuffling procedure was supposed to remove any energy-time correlation present in the data, the expected mean of the distribution is 0. The apparent deviation from 0 we interpret as the bias $\eta_\mathrm{bias}$ of our analysis. From the PDF we determine the $p$-value of the null hypothesis, i.e., the significance of the detection of a LIV effect, as its integral above $\eta_\mathrm{uncal.}$ and below $2\eta_\mathrm{bias}-\eta_\mathrm{uncal.}$. Our results for the theoretical LC, $p_{\eta_1}=0.78$ and $p_{\eta_2}=0.59$, are consistent with the null hypothesis.\\
\indent Note that this procedure is not applicable to the minimal LC model. Since the likelihood profile has no minimum, the bias is not well-defined. Furthermore, the minimal model is by construction valid only to obtain robust model-independent upper limits on $\eta$. 

\subsection{\label{sec:AppCI}Confidence interval calibration}
Since the PDF of $L$ is not a $\chi^2$ distribution, we cannot rely on the standard technique of finding the values of $\eta$ for which $L$ reaches the 3.84 threshold. Instead, we build a PDF of the values of $\eta$ corresponding to an arbitrary value of the $L$ threshold and calculate the quantiles of the PDF below (above) $\eta_\mathrm{bias}$. We repeat this procedure for different values of the $L$ threshold until the quantiles are 2.5\% (see Fig.~\ref{fig:Limits_distribution}). The value of the $L$ threshold obtained in this way we use to determine the ``uncalibrated'' upper (lower) limit $\eta^\mathrm{UL}_\mathrm{uncal.}$ ($\eta^\mathrm{LL}_\mathrm{uncal.}$). Finally, we compute the fully calibrated upper (lower) limits by subtracting $\eta_\mathrm{bias}$ from uncalibrated upper (lower) limits
\begin{align}
\eta^\mathrm{UL} = \eta^\mathrm{UL}_\mathrm{uncal.} - \eta_\mathrm{bias}\\ \eta^\mathrm{LL} = \eta^\mathrm{LL}_\mathrm{uncal.} - \eta_\mathrm{bias}.
\end{align}
\indent This procedure differs again from the standard Neyman construction of CIs, which is not feasible because it requires Monte Carlo simulations. However, it should produce equal results provided the PDF for $\eta-\eta_\mathrm{true}$ is symmetric with respect to its mean, and does not depend on $\eta_\mathrm{true}$. 

\end{document}